\documentclass[aps,pra,twocolumn,superscriptaddress,longbibliography]{revtex4-1}
\usepackage{braket}
\usepackage{graphicx}
\usepackage{amsmath}    
\usepackage{amsfonts}
\usepackage{xcolor}
\usepackage{multibib}

\newcites{SM}{SM References}

\begin{document}

\title{Sample-efficient adaptive calibration of quantum networks\\ using Bayesian optimization}

\author{Cristian L. Cortes}
\affiliation{Center for Nanoscale Materials,\\ Argonne National Laboratory, Lemont, Illinois 60439, USA}
\author{Pascal Lefebvre}
\affiliation{Institute for Quantum Science and Technology, and Department of Physics and Astronomy, University of Calgary, 2500 University Drive NW, Calgary, Alberta T2N 1N4, Canada}
\author{Nikolai Lauk}
\affiliation{Division of Physics, Mathematics and Astronomy, and Alliance for Quantum Technologies (AQT), California Institute of Technology, 1200 E. California Boulevard, Pasadena, California 91125, USA}
\author{Michael J. Davis}
\affiliation{Chemical Sciences and Engineering Division, Argonne National Laboratory, Lemont, Illinois 60439, United States}
\author{Neil Sinclair}
\affiliation{Division of Physics, Mathematics and Astronomy, and Alliance for Quantum Technologies (AQT), California Institute of Technology, 1200 E. California Boulevard, Pasadena, California 91125, USA}
\affiliation{John A. Paulson School of Engineering and Applied Sciences, Harvard University, 29 Oxford Street, Cambridge, Massachusetts 02138, USA}
\author{Stephen K. Gray}
\affiliation{Center for Nanoscale Materials,\\ Argonne National Laboratory, Lemont, Illinois 60439, USA}
\author{Daniel Oblak}
\affiliation{Institute for Quantum Science and Technology, and Department of Physics and Astronomy, University of Calgary, 2500 University Drive NW, Calgary, Alberta T2N 1N4, Canada}

\begin{abstract}
Indistinguishable photons are imperative for advanced quantum communication networks. 
Indistinguishability is difficult to obtain because of environment-induced photon transformations and loss imparted by communication channels, especially in noisy scenarios.
Strategies to mitigate these transformations often require hardware or software overhead that is restrictive (e.g. adding noise), infeasible (e.g. on a satellite), or time-consuming for deployed networks.
Here we propose and develop resource-efficient Bayesian optimization techniques to rapidly and adaptively calibrate the indistinguishability of individual photons for quantum networks using only information derived from their measurement.
To experimentally validate our approach, we demonstrate the optimization of Hong-Ou-Mandel interference between two photons--a central task in quantum networking-- finding rapid, efficient, and reliable convergence towards maximal photon indistinguishability in the presence of high loss and shot noise.
We expect our resource-optimized and experimentally friendly methodology will allow fast and reliable calibration of indistinguishable quanta, a necessary task in distributed quantum computing, communications, and sensing, as well as for fundamental investigations.
\end{abstract}

\maketitle

\section*{Introduction}
\begin{figure*}[t!]
    \center
    \includegraphics[width=0.95\textwidth]{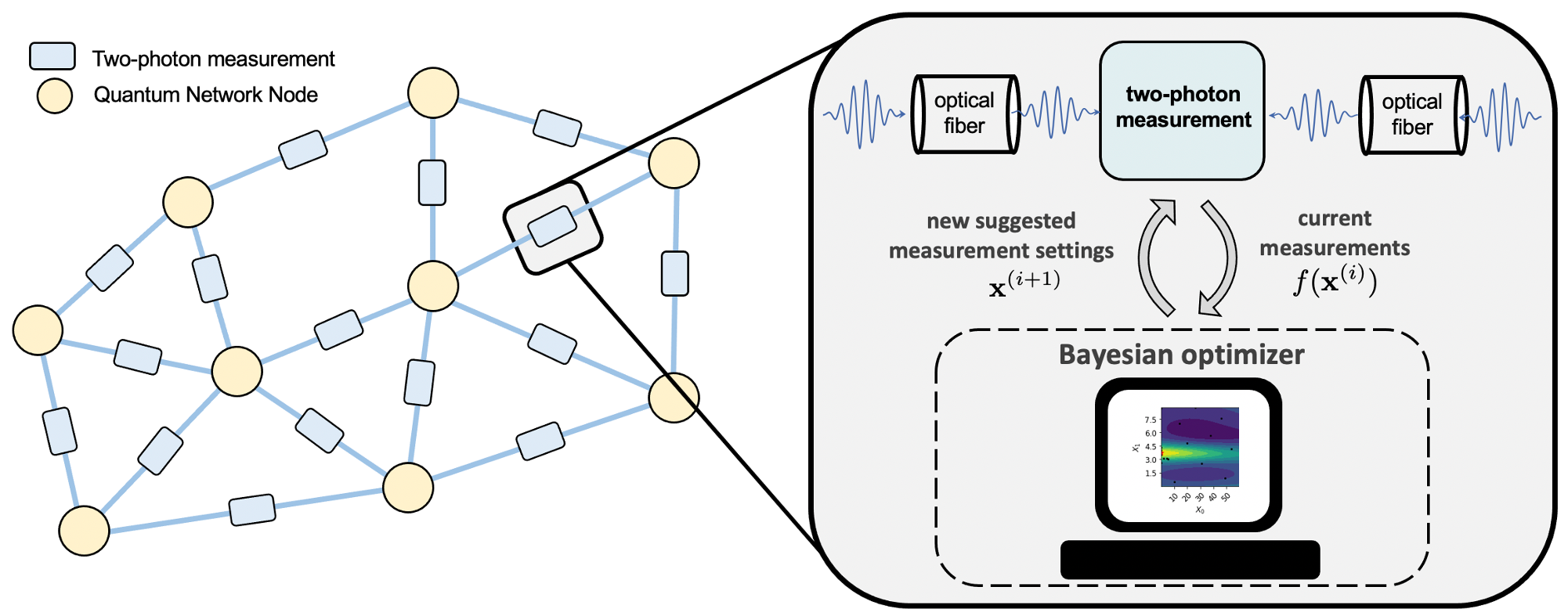}
    \caption{High-level overview of quantum network calibration using Bayesian optimization. Quantum network nodes emit single or entangled photons into fibers and two-photon Bell-state measurements, whose fidelities are determined by Hong-Ou-Mandel interference, are performed to facilitate quantum network protocols. 
    Within the two-photon measurement node, a feedback loop between the two-photon measurement apparatus and a Bayesian optimizer is used to automate the calibration of photon indistinguishability using the objective function  $f(\mathbf{x}^{(i)})$ which we introduce in the main text, and is plotted in the diagrammatic computer screen. The value $\mathbf{x}^{(i)}$ denotes the variable experimental degrees of freedom determined by the experimental apparatus during the $i$th iteration of the algorithm, with two variables of $\mathbf{x}$ defining the axes of the plot on the screen.
   }\label{fig:overview}
\end{figure*}

Algorithmic optimization of quantum systems plays a key role in quantum computing, simulation, and sensing (e.g. see \cite{carleo2017solving,havlivcek2019supervised,bogaerts2020programmable,kokail2019self,mott2017solving,carleo2017solving,peruzzo2014variational,ding2020retrieving,bonato2016optimized,lumino2018experimental,preskill2018quantum}), as well as for quantum system characterization \cite{gao2018experimental,tachella2019real,wittler2021integrated,simmerman2020efficient}. 
Yet, there has been little effort on algorithmic optimization of quantum communications and networks \cite{wallnofer2020machine,ismail2019integrating,melnikov2020setting,kottmann2021quantum}. 
In particular, to use such methods to overcome unavoidable channel-induced variations of properties (degrees of freedom) of photons, along with the well-known impacts of loss and noise, which restrict demonstrations of advanced, multi-qubit, quantum networks, especially those which crucially rely on interference \cite{kimble2008quantum,pirandola2015advances,moehring2007entanglement,delteil2016generation,pompili2021realization,bhaskar2020experimental,daiss2021quantum,guccione2020connecting,lucamarini2018overcoming,herbst2015entswap143km,ren2017ground,du2021elementary,valivarthi2016quantum,lago2021telecom,liu2021heralded}.
These variations, which originate from changes in the environment, render photons distinguishable, thereby restricting their ability to interfere \cite{gisin2002quantum}.
This precludes crucial tasks in a multi-node quantum network \cite{kimble2008quantum}, like that schematized in Fig.~\ref{fig:overview}, including two-photon Bell-state measurements \cite{michler1996expBSM} which underpin measurement-device-independent quantum key distribution (MDI-QKD) \cite{Lo2012} or quantum repeaters \cite{briegel1998quantum}, for example.
The required indistinguishability for deployed networks is obtained by calibrating all degrees of freedom of a photon, i.e. its polarization, temporal, spectral, and spatial modes.
This is a process that requires additional hardware and software, and relies on (often ``brute-force'') methods that either restrict the communication rate, add noise, do not scale to multi- photon or node networks, or are physically or financially infeasible for remote network nodes \cite{lucamarini2018overcoming,herbst2015entswap143km,ren2017ground,du2021elementary,valivarthi2016quantum,lago2021telecom,liu2021heralded}. 

Here we employ a Gaussian Process (GP) Bayesian optimization algorithm \cite{rasmussen2006} to rapidly calibrate the degrees of freedom of photons for quantum networks.
Our method operates with minimal resources: it requires only direct measurements of the low-rate streams of photons, which are inherent to quantum communications, with threshold detectors to overcome the impact of sampling noise and network channel-induced photon variations.
While Gaussian process modeling has been successfully used in many physical applications, for example to describe the optical response of plasmonic systems \cite{Miller2010,Miller2012}, its use in the quantum networking context, here for calibration, is an exciting new direction. 
Specifically, we present and experimentally demonstrate an adaptive calibration algorithm that maximizes two-photon Hong-Ou-Mandel (HOM) \cite{hong1987measurement} interference to efficiently render photons indistinguishable despite their low probability of detection (see Fig.~\ref{fig:overview}, right). 
Our proposed methodology for low-cost autonomous calibration leverages advantages of the Bayesian optimization framework in that it is model-agnostic, sample efficient, i.e., demonstrates convergence with minimal samples, and is robust to shot noise and, accordingly, is well-suited for the conditions of quantum communications.
We also test the Bayesian optimization algorithm with respect to the kernel, initial sampling strategy, and acquisition function to improve its robustness, and show that it provides the best performance compared to other approaches. We expect that our experimentally-friendly method will accelerate the development of high-fidelity quantum networks and reduce the complexity of implementing workable quantum technology.

\section*{Results}

\subsection*{Bayesian Optimization}
We envision using a GP Bayesian optimization algorithm \cite{rasmussen2006} to maximize the indistinguishability of photons generated at a quantum node and, after travelling through deployed fiber optics cable, arriving at a two-photon measurement station, as sketched on the right hand side of Fig.~\ref{fig:overview}.
In addition to being a powerful optimization framework, the final GP surrogate model allows for a deep analysis of a high-dimensional parameter space which, as described below, corresponds to degrees of freedom of the photons in this work. 
The calibration problem consists of optimizing an expensive objective function, $f(\mathbf{x})$, corresponding to a measure of the change in coincidence detections due to two-photon interference, as described in the next section. 
GP Bayesian optimization is based on the assumption that the likelihood and prior distributions correspond to multivariate Gaussian distributions, where the objective function is assumed to be a random variable sampled from a Gaussian Process,
\begin{equation}
    f(\mathbf{x}) \sim \mathcal{GP}[\mu(\mathbf{x}),K(\mathbf{x},\mathbf{x}')],
\end{equation}
which is defined by the mean function, $\mu(\mathbf{x}) = \mathbb{E}[f(\mathbf{x})]$, and covariance (or kernel) function, $K(\mathbf{x},\mathbf{x}') = \mathbb{E}[(f(\mathbf{x})-\mu(\mathbf{x})(f(\mathbf{x'})-\mu(\mathbf{x'}))]$ respectively. 
The Bayesian optimization algorithm proceeds by feeding the optimizer, a single function call which is equivalent to a single measurement result, $f(\mathbf{x}^{(i)})$, during the $i$th iteration of the algorithm,. 
The variable $\mathbf{x}^{(i)} = (x_1^{(i)},x_2^{(i)},\cdots,x_D^{(i)})$ is a $D$-dimensional vector corresponding to $D$ experimental parameters. 
For two-photon interference, these parameters control the photonic degrees of freedom that are adjusted to compensate for channel transformations. 
For our proof-of-principle laboratory demonstration described in the next section, these correspond to a time delay controlling the mutual arrival time of the photons, half-waveplate angle controlling the polarization of the photons, and  spectral filter pass-band determining the frequency detuning (offset) of the photons. 
Once the $i$th measurement result has been fed into the Bayesian optimizer, it builds a Gaussian Process surrogate model by performing an optimization with respect to kernel hyperparameters (see Appendix~\ref{appsec-bayesoverview}). 
An acquisition function, such as the lower confidence bound, which we found worked best in this work (see Appendix~\ref{appsec-acqfunct}), is then used to suggest the next measurement, $\mathbf{x}^{(i+1)}$. 
Various initial sampling strategies, kernel functions, and hyper-parameter optimization strategies are also tested with details of the algorithmic fine-tuning provided in Appendices~\ref{appsec-kernelfunct}-\ref{appsec-inisample}.

\begin{figure}[t] 
    \center
    \includegraphics[width=\columnwidth]{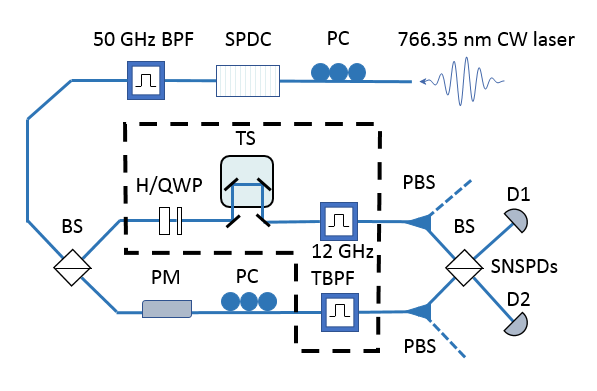}
    \caption{Experimental setup used for HOM interference. A diode laser generates 766.35~nm wavelength light that goes through a polarization controller (PC) before going through a periodically-poled lithium niobate crystal to generate a 1532.7~nm wavelength photon pair by Type-0 spontaneous parametric down conversion (SPDC). A 50 GHz-bandwidth bandpass filter (BPF) selects the degenerate photon pairs around 1532.7~nm wavelength and a beam splitter (BS), probabilistically separates each photon, directing them to different output paths. The top path, in free-space, directs the photon through half- and quarter-waveplates for control of the polarization, with the path length (i.e. time delay) controlled by a translation stage (TS), before the photon passes through a fiber-based polarizing beam splitter (PBS). The bottom path, entirely in fiber, has a phase modulator (PM) plus a polarization controller (PC) to maximize transmission through the PBS. The two paths meet back at a fiber-based beam splitter (BS) which performs HOM interference, and the photons are detected by superconducting nanowire single photon detectors (SNSPDs, identified as D1 and D2). Both paths also contain an optional 12~GHz-bandwidth tunable bandpass filter (TBPF). The components within the dashed line are those adjusted by the Bayesian optimizer.}
    \label{setup}
\end{figure}

\begin{figure*}[t]
    \centering
    \includegraphics[width=0.95\textwidth]{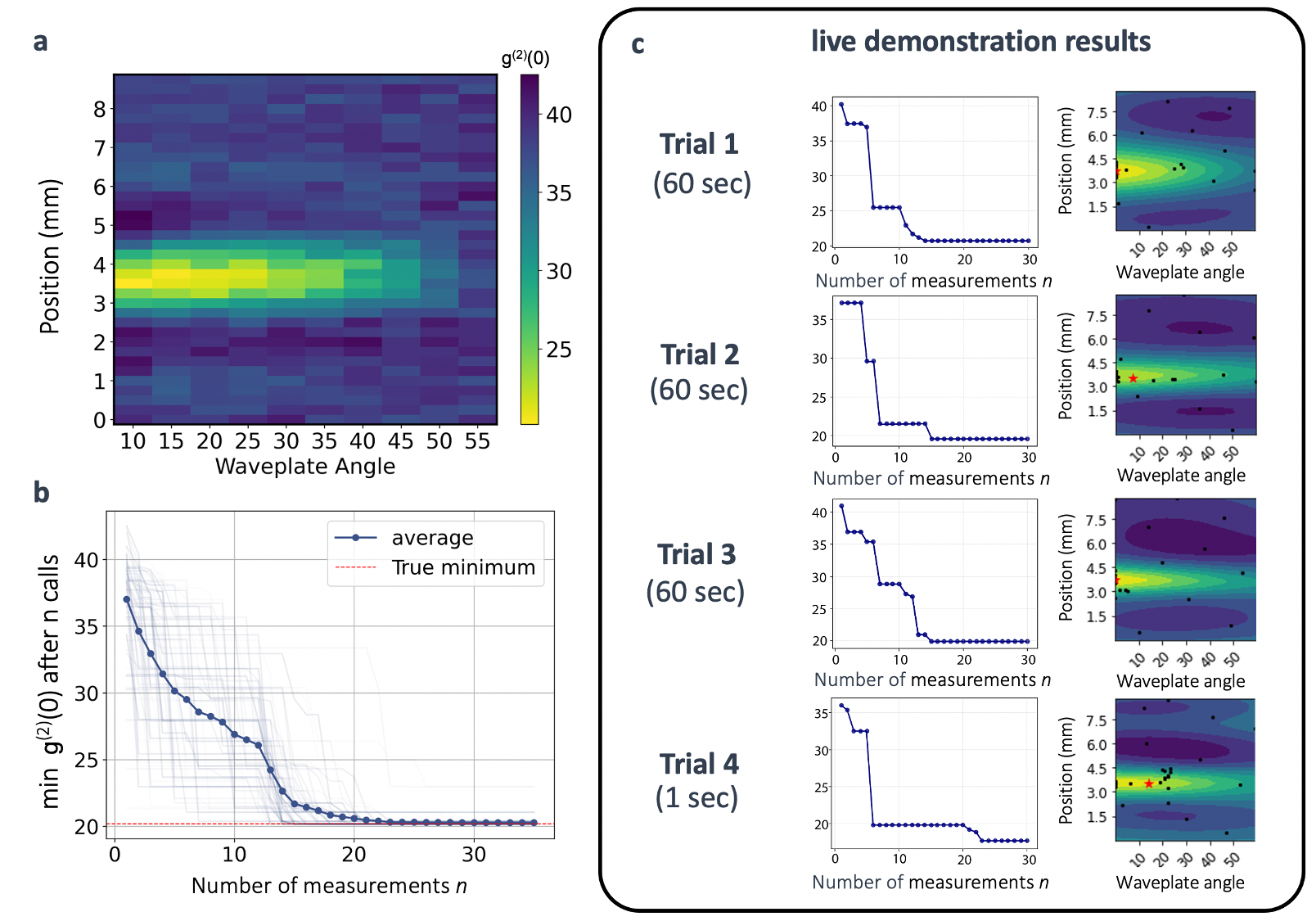}
    \caption{Real-time Bayesian optimization results by calibrating the photons using two degrees of freedom. (a) Normalized coincidence detections as a function of relative half-waveplate angle measured in degrees and relative translation stage position measured in millimeters (acquired over 13 h). This experimental data set is considered as the baseline function for all benchmark results. The results are symmetric about zero for negative relative polarization angles. (b) Full benchmark results using the baseline data set in (a) as a black-box function for the Bayesian optimizer. The dark blue line is the average benchmark result with respect to one hundred simulated trials. The true minimum corresponds to the minimum value from the baseline data in (a). (c) Live demonstration results using the Bayesian optimizer in four different trials with detector integration time in parentheses. Left: convergence plot. Right: Final surrogate model prediction. The black dots represent the sampled parameter settings and the red star shows the optimal settings predicted by the GP algorithm.}
    \label{map}
\end{figure*}

\subsection*{Experimental setup}
To test the algorithm, we performed a HOM interference measurement of two photons. 
When two indistinguishable independent photons are sent to different input ports of a beam splitter, they will bunch at one of the output ports \cite{hong1987measurement}. 
This results in a minimum (zero without imperfections) of the zero-delay-time normalized second-order correlation function \cite{Beck2007}, 
\begin{align}
g^{(2)}(0) = \frac{C_{12} T}{S_1 S_2 \Delta \tau}, 
\label{g2}
\end{align}
where $S_1$ and $S_2$ ($C_{12}$) denote the total number of photons detected at each (in coincidence at both) of the outputs of the beamsplitter within time interval $\Delta\tau$ over a total amount of time $T$. 
This quantity is related to the commonly used HOM interference visibility $V = (C_{12}^{\mathrm{max}}-C_{12}^{\mathrm{min}})/C_{12}^{\mathrm{max}}$, where $C_{12}^{\mathrm{max}}$ ($C_{12}^{\mathrm{min}}$) denotes the maximum (minimum) number of photons detected in coincidence as a degree of freedom of a photon is varied (e.g. its time of arrival) \cite{hong1987measurement,valivarthi2020teleportation}.
Either $V$ or $g^{(2)}(0)$ are used to quantify the impact of imperfections in HOM interference, which ultimately impacts Bell-state measurement fidelities, and hence the fidelity of qubit distribution in quantum networks \cite{nielsen2002quantum,metcalf2014quantum,valivarthi2020teleportation}.

In our experiment, which is depicted in Fig.~\ref{setup},a continuous-wave laser emits near-visible-wavelength light that pumps a periodically-poled lithium niobate crystal waveguide to create a photon pair at telecommunication wavelength through spontaneous parametric down conversion (SPDC) \cite{burnham1970observation}.
The leftover pump light is removed with a 50 GHz bandpass filter, and the photon pair is sent to an initial beam-splitter to probabilistically separate the photons.
Here we employ photons originating from the same source to demonstrate our principle even though photons would originate from independent sources in a quantum network.
To avoid first-order interference between correlated photons generated by the SPDC process, one of them passes through a fiber-based phase-modulator used to randomize the phase difference between the photon pair \cite{jin2013two}.
The second photon goes through a free-space optics setup in which a set of half- and quarter-waveplates adjust the photon polarization while mirrors on a translation stage vary the path length difference (corresponding to a mutual time delay) between each of the photons. 
Each photon then passes an optional, independent, and tunable 12 GHz-bandwidth bandpass filter.
The waveplates, translation stage, and tunable filters, surrounded by a dashed outline in Fig.~\ref{setup}, are adjusted according to the algorithm.
Next, each photon passes through fiber-based polarizing beam-splitters (PBSs), which ensure that the polarizations of both photons are identical and that any polarization rotations are converted into intensity variations.
Then, each photon is directed into a different input port of a second fiber-based beamsplitter at which two-photon interference occurs.
Finally, the output photons are guided to two cryogenically cooled superconducting nanowire threshold single-photon detectors. 
A time-to-digital converter records the time-of-arrival of the photons at the detectors.

\subsection*{Optimization using two degrees of freedom}
In our first measurement, we remove the tunable band-pass filters such that the path-length difference and the half-waveplate angle correspond to the experimental parameters $\mathbf{x}$ that are fed into the Bayesian optimization algorithm. 
It is important to point out that there are several choices for possible objective functions $f(\mathbf{x})$ one may use for the optimization algorithm. 
Perhaps unsurprisingly, we find that 
\begin{align}
f(\mathbf{x}) \equiv g^{(2)}(0)[\mathbf{x}]
\label{g2objective}
\end{align}
provides the best performance when compared to others and is what we employ for our scheme.
For example, $C_{12}$ can vary for experimental settings that do not yield quantum interference, or $V$ which requires adjustment of experimental parameters to assess both $C_{12}^{\mathrm{min}}$ and $C_{12}^{\mathrm{max}}$.
Note here that the zero in $g^{(2)}(0)$ indicates that we always measure about the same relative time-of-arrival of the photons (about $\Delta\tau=2$~ns) when we vary the degrees of freedom, specifically a time that corresponds to a minimum $g^{(2)}(0)$ when photons are rendered indistinguishable.

Before discussing the results of our optimization, note that we develop a theoretical model for $g^{(2)}(0)$ to validate our experimental measurements and the predictions of the Bayesian optimization algorithm (see Appendix~Fig.~\ref{fig:app-model}).
The degenerate output of our Type-0 SPDC crystal is described by a squeezed vacuum state, which is a Gaussian state, i.e. it is completely characterized by a displacement vector and a covariance matrix. 
Since all optical operations in the experiment, including the photon detection, can be described as Gaussian operations, that is, they map Gaussian states to other Gaussian states, we can apply a characteristic function formalism \cite{LaukInPrep, takeoka2015full} to determine the final displacement vector and the final covariance matrix, which allow us to predict $S_1$, $S_2$ and $C_{12}$. 
In this model, the experimental degrees of freedom are modeled as virtual beam splitters with variable transmittances $\eta_u,\eta_d$, and $\zeta$, where $\eta_{u/d}$ corresponds to overall photon coupling efficiency in the upper/lower path (see Fig.~\ref{setup}) and $\zeta=\exp(-x^2)$ corresponds to the mode overlap parameterization between the two photons. 
Details of the model are provided in the Appendix~\ref{appsec-HOMmodel} along with the theoretical plot of $g^{(2)}(0)$ in Appendix~Fig.~\ref{fig:app-g2model}a.

Our optimization using two degrees of freedom is depicted in Fig.~\ref{map}a, presenting a two-dimensional map of the measured $g^{(2)}(0)[\mathbf{x}]$ as a function of the full range of settings $\mathbf{x}$ of the translation stage and half-waveplate angle - using 60~s for changing the parameter settings and $T=60$ s of data-collection per setting. 
Our result shows good agreement with that predicted by the theoretical model as plotted in Appendix~Fig.~\ref{fig:app-g2model}a. 
In particular, $g^{(2)}(0)[\mathbf{x}]$ displays $V \leq 0.5$, as expected for a squeezed vacuum input state as the input to the initial beam splitter, which splits the pair probabilistically. 
Unlike the $g^{(2)}(0)[\mathbf{x}]$ predicted by the theoretical model, the experimental data exhibits random variations due to sampling noise.

To benchmark the Bayesian optimization algorithm, we used the baseline data  in Fig.~\ref{map}a as the source of values for our objective function $g^{(2)}(0)[\mathbf{x}]$ which are fed to the algorithm.
Starting with a set of twelve measurement settings $[\mathbf{x}^{(1)},\mathbf{x}^{(2)},\cdots,\mathbf{x}^{(12)}]$ selected based on Latin hypercube sampling (see Appendix~\ref{appsec-inisample} for details), the algorithm proceeds to select the next setting $\mathbf{x}^{(13)}$, and then the next and so on, for a total number of measurements, $n$, that it deems optimal for establishing the minimum of $g^{(2)}(0)[\mathbf{x}]$. 
As shown in Fig.~\ref{map}b, we find that the Bayesian optimization algorithm reliably converges to the minimum $g^{(2)}[\mathbf{x}]=20.2$ in less than $n=30$ measurements, on average, (dark blue line). 
The light blue lines correspond to single instances of the Bayesian optimization algorithm, which converge more slowly or more quickly depending on the random instance of initially sampled points.
By benchmarking the algorithm with over one-hundred independent trials, we are able to remove the effect of randomness, resulting in the average (expected) convergence shown by the dark blue line. 

While this benchmark confirms the validity of the approach, it is important to test the efficacy of our algorithm in a \emph{live} setting where every measurement is performed in real-time, as opposed to using the tabulated baseline data from Fig.~\ref{map}a. 
The results for four different real-time trials are shown in Fig.~\ref{map}c. 
The plots for trials 1-3, for which data was acquired for $T=60$~s (same as for the baseline data), show that the minimum value of $g^{(2)}(0)[\mathbf{x}]$ is found within $n=15$ in all three cases. 
The maps on the right of Fig.~\ref{map}c  show the final Gaussian Process surrogate model prediction made after $n=30$ measurements. 
It is worth noting that the surrogate models do not, and indeed are not intended to, accurately reproduce $g^{(2)}(0)[\mathbf{x}]$ over the entire parameter range of the input variables. 
The surrogate models need only accurately predict the location of the minimum of $g^{(2)}(0)[\mathbf{x}]$. 
In fact, this pursuit of incomplete but sufficient information is the underlying reason why  GP Bayesian optimization is more efficient than brute-force mapping of $g^{(2)}(0)[\mathbf{x}]$. 
In the fourth trial we reduce $T$ to just 1~s, which significantly increases the level of shot noise. 
Yet, the GP optimization still finds the minimum value after just $n=23$, albeit the polarization setting is slightly off the optimum value. 
As the PBS transmission follows a cosine of the half-waveplate angle $g^{(2)}(0)[\mathbf{x}]$ becomes less sensitive to small variations and, thus, renders the optimization more difficult. 
However, the consequence of a slight offset of the polarization angle is also less significant due to the cosine dependence of the loss. 
On the other hand, without the PBS the distinguishability would be directly proportional to the offset of the half-waveplate angle.

Our results using two degrees of freedom validate the effectiveness of the algorithm for a realistic signal that is subject to various sources of noise including shot noise, while demonstrating fast convergence. 
In comparison, the full experimental map (shown in Fig.~\ref{map}a) consists of 350 data points acquired over 13~h.

\begin{figure*}
    \centering
    \includegraphics[width=\textwidth]{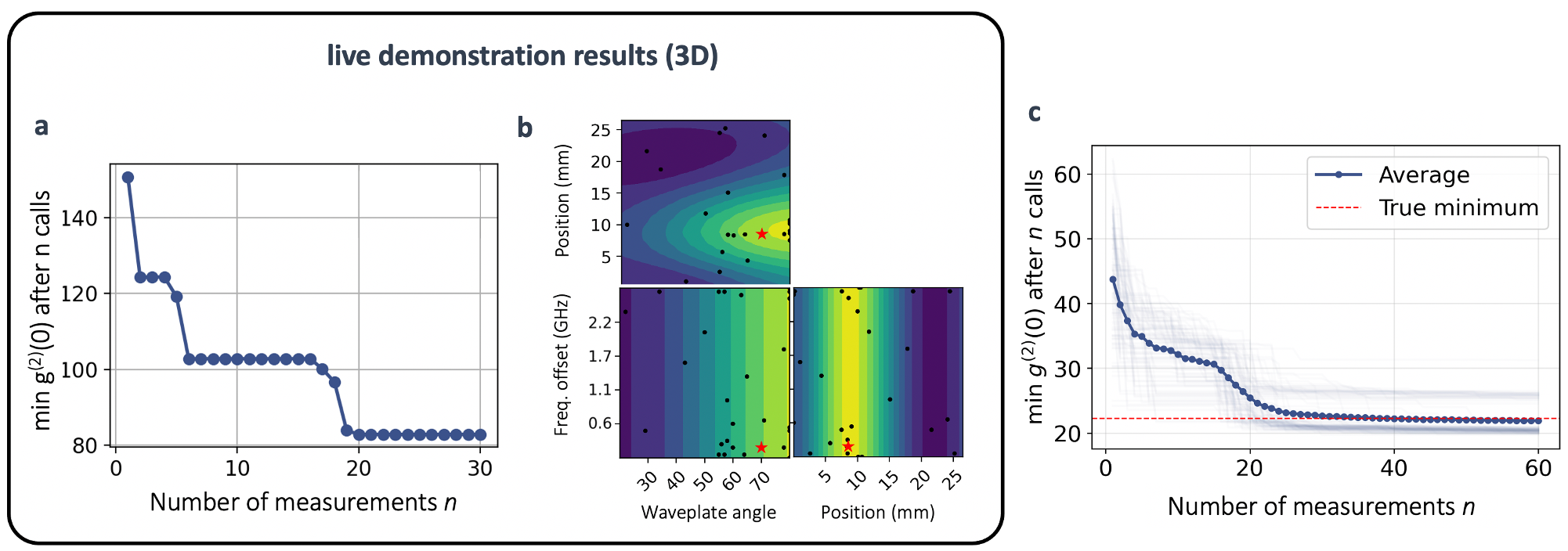}
    \caption{Real-time Bayesian optimization results using three degrees of freedom. (a) Convergence plot of the minimum $g^{(2)}(0)[\mathbf{x}]$ as a function of total number of measurements, $n$. (b) Partial dependence plots for each degree freedom (see main text for a description). (c) Benchmarking the optimization using a simulated objective function derived from the theoretical model. Note that theoretical model is based on same experimental parameters as those shown in Fig.~\ref{map}a, however, the additional spectral filtering for the three dimensional optimization lowers the mean-photon number and thus tends to increase $g^{(2)}(0)[\mathbf{x}]$.}
    \label{fig3D}
\end{figure*}

\subsection*{Optimization using three degrees of freedom}
Our next measurements test the Bayesian optimization algorithm in a higher-dimensional setting. 
Specifically, we add the 12 GHz-bandwidth tunable filters to the experimental setup so as to include the frequency offset (between 0 and 6 GHz) of the photons as a third degree of freedom.
At 0~GHz offset, both filters are resonant with the degenerate spectral mode of the photon pair, while at the maximum detuning of 6~GHz between the two photons, the spectral filters transmit correlated but partially non-degenerate photons. 
Compared to the previous setup, a broader region of interference with respect to varied path-length difference is expected due to the spectral filtering.
Since the parameter space of settings in three degrees of freedom is too large to acquire a baseline data-set as that in Fig.~\ref{map}a, we proceed directly to a live demonstration, with results shown in Fig.~\ref{fig3D}a. 
We find excellent convergence towards the minimum value within $n=25$ measurements. 
Two-dimensional partial-dependence plots of the thirty-point surrogate model are shown in Fig.~\ref{fig3D}b. 
The partial dependence plots illustrate the surrogate model prediction with respect to two degrees of freedom, with the remaining degree of freedom averaged out. 
We note that the first two degrees of freedom, corresponding to the half-waveplate angle and path length difference, display a near minimum at around 85~degrees and 7.5~mm respectively, as expected. 
The final degree of freedom, corresponding to the frequency offset, displays a near-constant dependence with a small slope predicting a minimum at $X_2=0$ (zero-frequency offset). 

Using our theoretical model, we generate a map of the predicted $g^{(2)}(0)$ for all possible parameter settings. 
Note that our model captures the spectral offset of the two photons as a transition between degenerate squeezed vacuum to non-degenerate two-mode squeezed vacuum states as inputs to the final beamsplitter, as shown in Appendix~Fig.~\ref{fig:app-g2model}a,b.
Thus, our theoretical model, instead of an experimentally acquired baseline map, is used as the source of the $g^{(2)}(0)[\mathbf{x}]$ values employed for systematically benchmarking the Bayesian optimization algorithm for the three-dimensional parameter set.
In Fig.~\ref{fig3D}c we plot the value of the predicted $g^{(2)}(0)[\mathbf{x}]$ as a function of the total number of measurements $n$ averaged over a hundred trials. 
A set of fifteen measurement settings based on Latin hypercube sampling are used as the initial sampling points. 
The results show excellent convergence towards the predicted minimum within $n=40$ measurement calls, showing the efficacy of the result in the higher dimensional setting while also demonstrating the utility of the theoretical model.

\subsection*{Optimization for simulated thermal input states}

\begin{figure*}
    \centering
    \includegraphics[width=\textwidth]{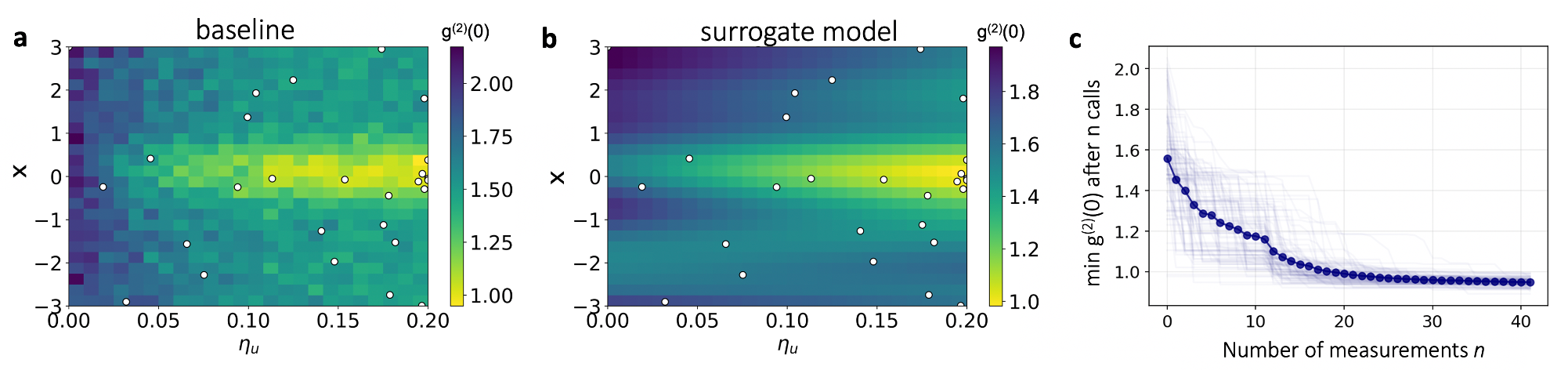}
    \caption{Simulation results for Bayesian optimization using thermal input states. (a) Simulated baseline data using Poisson sampling. (b) Final surrogate model prediction after 40 measurement calls. (c) Convergence plot averaged over 150 instances (dark blue).}
    \label{thermal}
\end{figure*}

For deployed quantum networks, where photon sources are separated, one will not interfere correlated photons as in our demonstration. 
Instead, two-photon measurements are typically performed on two uncorrelated coherent photons, as in the case of MDI-QKD \cite{Lo2012}, or two photons, each entangled with another photon not partaking in the measurement, as in the case of quantum repeaters \cite{briegel1998quantum}. 
For the latter case, photons from separate SPDC sources will be independently calibrated in order to maximize the degree of photon indistinguishability. 
When one photon from an SPDC source is observed without its correlated partner, it will follow a thermal photon number distribution \cite{burnham1970observation}. 
Hence, we provide additional benchmarks with thermal states at the input of the beamsplitter -- corresponding to two independent members of separate photon pairs -- using our theoretical model for the predicted $g^{(2)}(0)$, see Appendix~Fig.~\ref{fig:app-g2model}c.
We simulate the effect of finite sampling by performing Poisson sampling of the theoretical model, which we define as the baseline data. The independent degrees of freedom correspond to $\zeta=\exp(-x^2)$ (due to path length difference) and $\eta_u$ (loss due to polarization misalignment), which in an experimental setting would be proportional to the cosine of the polarization, see the Appendix~\ref{appsec-HOMmodel} for more details. 
The result is shown in Fig.~\ref{thermal}a. 
The effect of finite sampling is evident by the random noise, which reflects what a real acquired signal would look like, e.g. as in Fig.~\ref{map}a. 
The baseline objective function is, thus, used to test the Bayesian optimizer under experimental conditions for which the photon detection rate can be very limited. 
The results shown in Fig.~\ref{thermal} provide the final benchmark results for Bayesian optimization using thermal input states, averaged with 150 different instances. 
As before, a set of twelve measurement settings based on Latin hypercube sampling are used as the initial sampling points. 
The results shown in Fig.~\ref{thermal}c show that the Bayesian optimizer finds the expected minimum within $n=30$ measurements (located near $\eta_u = 0.2$ and $x=0$). 
These results are promising for in situ quantum network calibration.

\section*{Discussion}

It is worth discussing how another widespread optimization approach, gradient-based optimization methods, would operate and compare to the Bayesian optimization framework presented in the manuscript. 
First-order gradient descent based methods, such as conjugate gradient descent, find the local or global minimum by using the update rule, $\mathbf{x}^{(i+1)} = \mathbf{x}^{(i)} - \gamma \nabla f(\mathbf{x}^{(i)})$, where $\gamma$ is the step size value and $\nabla f(\mathbf{x}^{(i)})$ corresponds to the gradient of the objective function. 
While gradient descent scales well to high dimensions, it becomes very time consuming for expensive functions where gradient information is not available. 
In principle, it is possible to obtain gradient information for the current set-up, but it would require an additional measurement at each point in order to use the finite-difference formula, $\tfrac{\partial f}{\partial\theta}|_a = \lim_{\delta\theta\rightarrow 0} (f(a+\delta\theta)-f(a))/\delta\theta$, resulting in a doubling of the total number of measurements. 
In addition, vanilla gradient descent methods typically converge with hundreds of function calls at a minimum, therefore, they would be much more time-consuming than the Bayesian optimization presented here. Similar reasoning applies to second-order methods which require second-order derivatives to construct the Hessian. While second-order methods converge much faster than first-order methods, they come at the expense of additional function calls to construct the Hessian at each optimization step. 

The computational bottleneck of Bayesian optimization is also an important consideration for high-dimensional problems. The computational cost of the inference step in Bayesian optimization involves solving the linear system of equations, $(K+\sigma^2 I)^{-1}\mathbf{y}$, which scales as $\mathcal{O}(n^3)$ with $\mathcal{O}(n^2)$ storage, where $n$ is equal to the total number of sampling points. The predictive mean and variance prediction scales as $\mathcal{O}(n)$ and $\mathcal{O}(n^2)$ respectively per test point. This implies that for a large number of iterations, the computational cost can become prohibitive. To reduce computational complexity, many approximation schemes have been proposed \cite{wilson2015kernel}. For example, it may be possible to use the KISS-GP approach of Wilson and Nickisch \cite{wilson2015kernel} which reduces the inference complexity to $\mathcal{O}(n)$ and the test point prediction complexity to $\mathcal{O}(1)$. This approach decomposes covariance matrices in terms of Kronecker products of Toeplitz matrices, which naturally occur in 1D regularly spaced grids, which are then approximated by circulant matrices. By performing local kernel interpolation, it then becomes possible to speed up Bayesian optimization resulting in the much improved computational cost mentioned above. Other methods such as covariance matrix decomposition, likelihood approximations, penalized likelihood, low-rank approximations, and graphical representation may also be used for scalability and will be explored in future work \cite{geoga2020scalable,stein2004approximating,bilionis2013multi,krock2021modeling,constantinescu2013physics,constantinescu2013physics,constantinescu2006autoregressive}.

The advantage we have demonstrated in using Bayesian optimization for two-photon interference-based for quantum communications is for the case where the parameters that we adjust are mutually independent.
However, our demonstration could be extended to more complex scenarios, like linear quantum repeaters based on two-photon interference and entanglement swapping, which may not allow for mutually independent adjustment, depending on the approach.
For instance, the emission wavelength of light from a laser may drift, and will need to be adjusted. 
Thus, if photon pairs are created using SPDC, such adjustment will vary the wavelengths of both photons (or reduce the distribution rate, depending on the filters used).
More crucially, the use of entangled states fundamentally links the properties of two photons together by way of the relative phase of the qubit bases \cite{franson1989bell}, a property that is not considered in our demonstration, and one that is also relevant for single photon-based repeaters \cite{cabrillo1999creation} or twin-field QKD \cite{lucamarini2018overcoming}.
For instance, the wavelength and phase of photons in an entangled pair are coupled, and thus an optimization algorithm will need to account for the increased complexity when more links are employed to reach greater distances, as well as for calibration of the measurement bases at the end of the communication channel (e.g. the relative phase of an interferometer when using time-bin encoding).
Investigation of our, as well as other optimization methods, including its scaling advantage with communication distance, and how it applies to other network protocols, topologies, or encodings (e.g. all-photonic repeaters \cite{azuma2015all}, larger Hilbert spaces \cite{bergmann2019hybrid}, or continuous-variables \cite{pirandola2015advances}), or other quantum tasks \cite{nielsen2002quantum} will be considered in future work.

\section*{Conclusion}
We have presented and demonstrated a GP Bayesian optimization framework for both accelerating and automating the calibration of photonic degrees of freedom in order to maximize photon indistinguishability, which is an inescapable task in future quantum networks. 
The methodology is sample-efficient, easy-to-use, and has small computational overhead for a small number of optimization dimensions (ranging from two to ten). 
It is also robust to noise and experimental imperfections, making it a suitable as a plug-and-play approach for calibrating quantum network experiments from scratch. We have also implemented a theoretical model based on Gaussian operations to validate our optimization and generate baseline data. 
We envision that our optimization approach should be applicable to a wide variety of quantum optics experiments outside of the quantum network experimental focus for the current manuscript. 
For example, this approach could be useful for calibrating quantum optical computational devices \cite{metcalf2014quantum,qiang2018large,spring2013boson,arrazola2021quantum}, in particular in distributed computing architectures, as well as quantum imaging and spectroscopic methods \cite{brida2010experimental,lemos2014quantum}, especially in prototype systems where loss and noise play a large role.
In particular, as quantum networks become more complex and with greater demands on performance, efficient and practical optimization methods must be considered to overcome the deleterious impact of the environment on qubit transmission rates and fidelities.

\section*{Acknowledgements}
We thank E. M. Constantintescu for discussions on Gaussian process modeling and R. Valivarthi for discussions in the early stages of this work.
We thank Jean-Roch Vlimant for reading and commenting on the manuscript.
The detectors were provided by V. B. Verma, F. Marsili, M. Shaw, and S.W. Nam. 
The SPDC waveguides were provided by L. Oesterling through a collaboration with W. Tittel. 
This work was performed, in part, at the Center for Nanoscale Materials, a U.S. Department of Energy Office of Science User Facility, and supported by the U.S. Department of Energy, Office of Science, under Contract No. DE-AC02-06CH11357. 
This work is supported by the U. S. Department of Energy, Office of Basic Energy Sciences, Division of Chemical Sciences, Geosciences, and Biosciences operating under Contract Number DE-AC02-06CH11357.
N.L. and N.S. acknowledge support from the AQT Intelligent Quantum Networks and Technologies (INQNET) research program and DOE/HEP QuantISED program grant, QCCFP/Quantum Machine Learning and Quantum Computation Frameworks (QCCFP-QMLQCF) for HEP, Grant No. DE-SC0019219. 
N.S. further acknowledges the support of the Natural Sciences and Engineering Research Council of Canada (NSERC). 
P.L. and D.O. acknowledge support of the Natural Sciences and Engineering Research Council of Canada through the Discovery Grant program and the CREATE QUANTA training program, and the Alberta Ministry for Jobs, Economy and Innovation through the Major Innovation Fund Quantum Technologies Project (QMP).

\bibliography{REFERENCES}

\newpage

\setcounter{equation}{0}
\renewcommand{\theequation}{A.\arabic{equation}}
\setcounter{figure}{0}

\renewcommand{\figurename}{Appendix Fig.}

\section*{Appendix}
\subsection{Bayesian optimization overview}\label{appsec-bayesoverview}

Bayesian optimization is a sample-efficient technique that performs sequential optimization of time-consuming black-box functions. There are two main steps to any Bayesian optimization algorithm: (1) the construction of a conditional probabilistic model for the objective function based on a set of observations, and (2) the construction of an acquisition function that uses this model to predict future observations that optimize the objective function of interest. The first step requires an understanding of Bayesian inference, which quantifies how we can update our belief of a particular hypothesis of the objective function based on the current set of observations. The second step requires defining an effective criterion that may be used to predict new observation points that will (most likely) be close to the optimum. In the following section, we will introduce basic concepts related to Bayesian inference.  The second section will define Gaussian Processes (GP) which are often used as computationally tractable surrogate models for Bayesian optimization. The third section will present various kernel functions relevant to Bayesian optimization. The fourth section will define and discuss various acquisition functions which are often used in practice. The fifth section will define the quantum network black-box objective used for this study.  

\subsection{Bayesian Inference}\label{appsec-bayesinfer}

Bayesian inference aims to construct a conditional probability distribution $p(\mathcal{H}|\mathcal{D})$ for a hypothesis $\mathcal{H}$ based on observed data $\mathcal{D}$. For our purposes, the hypothesis represents the real values of an objective function $f(\mathbf{x})$ where $\mathbf{x}$ is a set of experimental knobs/parameters within the quantum optical experiment. In the context of quantum networks, we are interested in maximizing photon indistinguishability in order to maximize the performance of a wide variety quantum network operations and protocols. A single-measurement objective function  which is able to quantify the degree of photon indistinguishability is the normalized second order photon correlation function, $f(\mathbf{x}) \equiv g^{(2)}(0)[\mathbf{x}]$, measured with the Hong-Ou-Mandel experimental set-up shown in the main text. Here, the goal is to minimize the second order correlation function in order to maximize indistinguishability. For general Bayesian optimization, the experimental knobs can be categorical, integer, or real-valued quantities but for our purposes, we will only consider real-valued quantities such as the polarization rotation angle, position delay stage, and frequency filter bandwidth. 

In the following, we  define the basic concepts of Bayesian inference without reference to the photon correlation function, however, we will later refine the likelihood functions and kernel functions for the Gaussian Process in order to refine the Bayesian optimization algorithm to the quantum network problem. 

\subsubsection*{Bayes Theorem}

Let $X = [\mathbf{x}^{(1)},\cdots,\mathbf{x}^{(n)}]$ represent the matrix of observed input vectors $\mathbf{x}^{(i)} = (x_1^{(i)},x_2^{(i)},\cdots,x_D^{(i)})$ defined in a $D$-dimensional space, and $\mathbf{y} = [y^{(1)},\cdots,y^{(n)}]^T$ represent the vector of observed outputs $y^{(i)}$. The posterior distribution $p(\mathcal{H}|\mathcal{D})$ conditioned on the observed data $\mathcal{D}=\{X,\mathbf{y}\}$ is computed using Bayes' theorem,
\begin{equation}
    p(\mathcal{H}|\mathcal{D}) = \frac{p(\mathcal{D}|\mathcal{H})p(\mathcal{H})}{p(\mathcal{D})},
\end{equation}
where $p(\mathcal{D}|\mathcal{H})$ is the likelihood function and $p(\mathcal{H})$ is the prior probability distribution for the hypothesis $\mathcal{H}$. Bayesian inference uses Bayes' theorem (1) to update the probability of our hypothesis $\mathcal{H}$ as more information becomes available. The Bayesian approach provides several advantages including:  (i) providing a full probabilistic description of our hypothesis (such as parameter estimates) rather than point estimates, (ii) it is generally more robust to noise and outliers, (iii) it allows for inclusion of prior knowledge, (iv) and it is straightforward to use in the small sample size limit. For a given posterior distribution, the posterior predictive distribution for a new data point $\mathcal{D}^*=\{\mathbf{x}^*,y^*\}$ is defined as:
\begin{equation}
    p(\mathcal{D}^*|\mathcal{D}) = \int p(\mathcal{D}^*|\mathcal{H})p(\mathcal{H}|\mathcal{D})d\mathcal{H}.
\end{equation}
This quantity predicts the distribution of new, unobserved data. Our hypothesis and prior knowledge will control the convergence of the Bayesian optimization algorithm. We will discuss the Gaussian Process priors below. See Ref. \onlinecite{rasmussen2006} for a more detailed discussion of Bayesian optimization and Gaussian processes. 

\subsection{\textbf{Gaussian Processes}}\label{appsec-gaussproc}
For arbitrary likelihood and prior distributions, the exact calculation of the posterior distribution is not tractable. In particular, the denominator, which acts as a normalization constant for the posterior to remain a valid probability distribution, requires the calculation of a high-dimensional integral that is generally intractable. It is possible to use Monte-Carlo-based methods, however, these approaches are generally computationally expensive and ultimately become intractable for high dimensions. By assuming that the likelihood and prior correspond to multivariate Gaussian distribution functions, it is possible to calculate the posterior and predictive posterior exactly. In the Gaussian Process framework, we assume that the training and test data have Gaussian noise and may be written in the form, 
\begin{equation}
    y^{(i)} = f(\mathbf{x}^{(i)}) + \epsilon^{(i)},
\end{equation}
where $\epsilon_i$ corresponds to Gaussian-distributed noise with zero mean, $p(\epsilon) = \mathcal{N}(0,\sigma^2)$. The likelihood of the observed data $\mathcal{D}$ is then given by:
\begin{equation}
    p(\mathbf{y}|X,\mathbf{f}) = \mathcal{N}(\mathbf{f},\sigma^2I).
    \label{likelihood}
\end{equation}
The ``noiseless'' signal $\mathbf{f} = [f(\mathbf{x}^{(1)},f(\mathbf{x}^{(2)}),\cdots,f(\mathbf{x}^{(n)})]^T$ plays the role of the hypothesis $\mathcal{H}$ which corresponds to the parameters we wish to estimate. In this regard, the Gaussian Process approach is model-agnostic, which may be advantageous in scenarios where parametric models are not available, or perhaps undesirable to avoid the introduction of model bias. Since the noiseless vector $\mathbf{f}$ plays the role of the effective parameters, we require defining a prior distribution on these parameters. The Gaussian Process framework assumes the prior distribution is a Gaussian Process
written as, $f(\mathbf{x}) \sim \mathcal{GP}(\mu(\mathbf{x}),k(\mathbf{x},\mathbf{x}'))$, where the symbol ``$\sim$'' is read as: ``\emph{is sampled from}''  or ``\emph{is distributed as}.'' This implies that we treat the measured outputs $f_i$ as random variables that are sampled from the Gaussian process distribution function. Note that a Gaussian process is a distribution over functions completely defined by the mean function $\mu(\mathbf{x}) = \mathbb{E}[f(\mathbf{x})]$ and covariance function $k(\mathbf{x},\mathbf{x}') = \mathbb{E}[(f(\mathbf{x})-\mu(\mathbf{x}))(f(\mathbf{x}')-\mu(\mathbf{x}'))]$. The prior distribution for $\mathbf{f}$ can therefore be written as a multivariate Gaussian, 
\begin{equation}
    p(\mathbf{f}) = \mathcal{N}(\mu(\mathbf{x}),K)
    \label{prior}
\end{equation}
where $K$ is the kernel covariance matrix,
\begin{align}
    K &\equiv K(X,X) \\
    &=\begin{bmatrix}
    k(\mathbf{x}^{(1)},\mathbf{x}^{(1)}) & k(\mathbf{x}^{(1)},\mathbf{x}^{(2)}) & \cdots & k(\mathbf{x}^{(1)},\mathbf{x}^{(n)}) \\
    k(\mathbf{x}^{(2)},\mathbf{x}^{(1)}) & k(\mathbf{x}^{(2)},\mathbf{x}^{(2)}) & \cdots & k(\mathbf{x}^{(2)},\mathbf{x}^{(n)}) \\
    \cdots & \cdots  & \cdots  & \cdots \\
    k(\mathbf{x}^{(n)},\mathbf{x}^{(1)}) & k(\mathbf{x}^{(n)},\mathbf{x}^{(2)}) & \cdots & k(\mathbf{x}^{(n)},\mathbf{x}^{(n)})
    \end{bmatrix}, \nonumber
\end{align}
and must be a positive semi-definite matrix. Given the likelihood (\ref{likelihood}) and prior distribution (\ref{prior}), it is possible to calculate the posterior as:
\begin{align}
    p(\mathbf{f}|X,\mathbf{y}) &= \frac{p(\mathbf{y}|X,\mathbf{f})p(\mathbf{f})}{p(\mathbf{y}|X)} \\
    &=\mathcal{N}( K(K + \sigma^2 I)^{-1}\mathbf{y}, \sigma^2(K + \sigma^2 I)^{-1} K) \nonumber
\end{align}
where, as an aside, it is worth noting that $\sigma^2(K + \sigma^2 I)^{-1} K = K - K(K + \sigma^2 I)^{-1} K$, and the latter quantity is often found in textbooks. Let us now calculate the predictive distribution for unobserved data, $\mathcal{D}^* =\{X^*,\mathbf{y}^*\}$. The joint Gaussian process prior for the observed and unobserved data is written as:
\begin{equation}
    p
    \begin{pmatrix}
    \mathbf{f} \\
    \mathbf{f}^*
    \end{pmatrix}
    =
    \mathcal{N}\left[
    \begin{pmatrix}
    \mu(\mathbf{x})\\
    \mu(\mathbf{x}^*)
    \end{pmatrix},
    \begin{pmatrix}
    K_{x,x} + \sigma^2 I & K_{x,x^*} \\
    K_{x^*,x} & K_{x^*,x^*}
    \end{pmatrix}
    \right].
\end{equation}
The predictive distribution, $p(\mathbf{f}^*|\mathcal{D}) = \int p(\mathbf{f}^*|\mathbf{f})p(\mathbf{f}|\mathcal{D}) d\mathbf{f}$, is then found to be:
\begin{equation}
    p(\mathbf{f}^*|\mathcal{D}) = \mathcal{N}( \mathbf{f}^*| \mu_p(\mathbf{x}^*),\Sigma_p(\mathbf{x}^*) )
\end{equation}
with $\mu_p(\mathbf{x}^*)$ and $\Sigma_p(\mathbf{x}^*)$ corresponding to the predicted mean and variance of the model at point $\mathbf{x}^*$,
\begin{align}
    \mu_p(\mathbf{x}^*) &= \mu(\mathbf{x}^*) + k_*^T(K + \sigma^2 I )^{-1}(\mathbf{f}-\mu(\mathbf{x}) ) \label{key1}\\
    \Sigma_p(\mathbf{x}^*) &= k(\mathbf{x}^*,\mathbf{x}^*) - k_*^T(K + \sigma^2 I )^{-1}k_* \label{key2}
\end{align}
where $k_* = k(\mathbf{x},\mathbf{x}^*)$ is an $(n\times m)$ matrix corresponding to the $n$ training points $\mathbf{x}$ and $m$ test points $\mathbf{x}^*$. Equations (\ref{key1}) and (\ref{key2}) are the key equations which describe Gaussian-Process-based Bayesian inference.

\subsection{\textbf{Kernel functions}}\label{appsec-kernelfunct}
The covariance function encodes information about the shape and structure the objective function. The kernel ultimately affects the convergence rate of the Bayesian optimization algorithm. Below, we write several well-known kernel functions from the literature. For example, the squared exponential kernel is written as:
\begin{equation}
    k_(\mathbf{x},\mathbf{x}') = \exp\left( -\frac{||\mathbf{x} - \mathbf{x}'||^2}{2\ell^2} \right),
\end{equation}
which ensures that nearby points have similar function values within the length scale given by $\ell$. The periodic kernel is given by:
\begin{equation}
    k (\mathbf{x},\mathbf{x}') = \exp\left( -\frac{2 \sin^2(\pi ||\mathbf{x} - \mathbf{x}'||/p)}{\ell^2} \right)
\end{equation}
where $\ell$ is a length scale parameter and $p$ is the periodicity of the kernel. Finally, we also consider the Matern kernel,
\begin{align}
    k (\mathbf{x},\mathbf{x}') = \frac{1}{\Gamma(\nu)2^{\nu-1}}\left( \tfrac{\sqrt{2\nu}}{l} ||\mathbf{x} - \mathbf{x}'||^2\right)^\nu \!\!
    K_\nu( \tfrac{\sqrt{2\nu}}{l} ||\mathbf{x} - \mathbf{x}'||^2),
 \end{align}
 where $K_\nu$ is a modified Bessel function and $\Gamma$ is the gamma function. The parameter $\nu$ controls the smoothness of the function. The smaller the $\nu$, the less smooth the function will be. A comparison of various kernels is shown in Appendix~Fig.~\ref{fig:app-kernel}, demonstrating that the Matern kernel provides the best performance compared to all other kernels.  

\subsubsection*{Learning the hyperparameters of the kernel}
Given data set $\mathcal{D}$, the kernel hyperparameters such as $\ell$ and $p$ (here, we denote them as $\theta$) will strongly affect the final GP surrogate model predictions. How do we choose the correct hyperparameters values that should be used for the posterior $p(\mathbf{f}|\mathcal{D})$? The idea here is to calculate the probability of observing the given data under our prior: $p(\mathbf{y}|X,\theta) = \int p(\mathbf{y}|\mathbf{f})p(\mathbf{f}|X,\theta) d\mathbf{f}$, which is referred to as the marginal likelihood. The hyperparameters $\theta$ can be determined by performing maximum likelihood estimation of the marginal likelihood $p(\mathbf{y}|\mathbf{x},\theta)  =  \mathcal{N}(\mu(\mathbf{x}),k_\theta(\mathbf{x},\mathbf{x}')+\sigma^2 I)$. Taking the logarithm of this function, we obtain
\begin{align}
    \ln p(\mathbf{y}|\mathbf{x},\theta) &= -\frac{1}{2}\ln\det(k_\theta(\mathbf{x},\mathbf{x}')+\sigma^2 I)      \label{loglikelihood}
\\
    &- \tfrac{1}{2}(\mathbf{y} - \mathbf{\mu})^T(k_\theta(\mathbf{x},\mathbf{x}')+\sigma^2 I))^{-1}(\mathbf{y} - \mathbf{\mu}), \nonumber
\end{align}
where we have dropped terms that are not dependent on the kernel hyperparameters $\theta$. Here, the first term corresponds to the volume of the prior and becomes large when the volume of the prior is small (i.e. when the model is simple), while the second term becomes large when the data fits the model very well. Assuming that the posterior distribution over hyperparameters $\theta$ is \emph{well-concentrated}, we can approximate the predictive posterior as
\begin{equation}
    p(\mathbf{f}^*|\mathcal{D}) \approx p(\mathbf{f}^*|\mathcal{D},\theta_{MLE})
\end{equation}
where $\theta_{MLE}$ corresponds to the hyperparameters that maximize the log-likelihood (\ref{loglikelihood}),
\begin{equation}
    \theta_{MLE} = \text{argmax}_\theta \;p(\mathbf{y}|\mathbf{x},\theta).
\end{equation}

\begin{figure}[h!]
    \centering
    \includegraphics[width=\columnwidth]{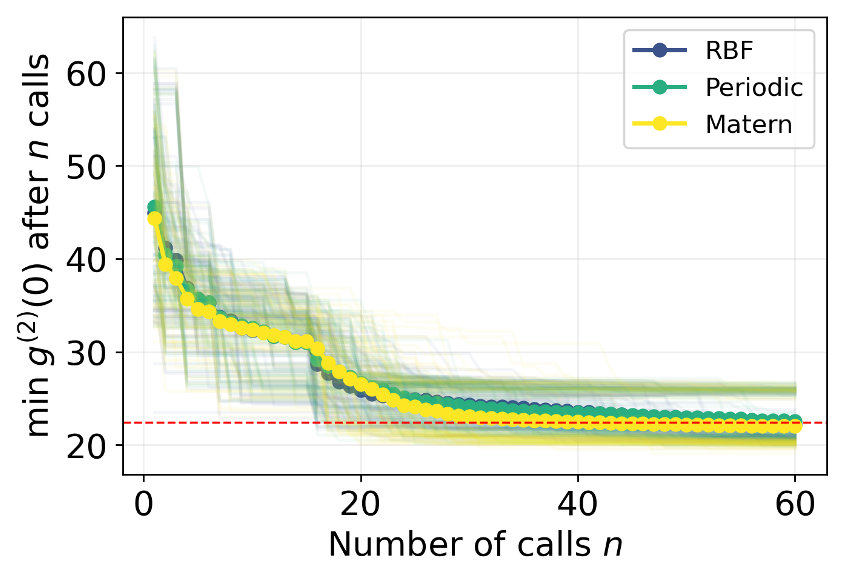}
    \caption{Convergence plots for different kernels including the radial basis function RBF kernel (also known as squared exponential kernel), the periodic kernel, and the Matern kernel ($\nu=5/2$) using the 3D baseline model for testing purposes.}
    \label{fig:app-kernel}
\end{figure}

\subsection{\textbf{Acquisition Functions}}\label{appsec-acqfunct}
An acquisition function $a(x)$ aims to evaluate the expected loss associated with evaluating $f(\mathbf{x})$ at point $\mathbf{x}$, and selects the point with the lowest \emph{expected} loss. In the following, we compare three different acquisition functions which are often used in the literature.

\subsubsection*{(i) Probability of improvement}
\noindent
Let $f'$ denote the minimum value of the objective function that has been observed so far. The probability of improvement aims to evaluate $f$ at the location most likely to improve upon this value. Here, we define the utility function,
\begin{equation}
    u(x) = 
    \begin{cases}
    0 & f(x) > f'\\
    1 & f(x) \leq f'.
    \end{cases}
\end{equation}
This utility function implies that a unit reward is given if $f(x)$ is less than $f'$, and provides no reward otherwise. The probability of improvement is the expected utility as a function of x:
\begin{align}
    a_{PI}(x^*) = \mathbb{E}\left[ u(x^*)| x^*,\mathcal{D} \right] &= \int_{-\infty}^{f'} \mathcal{N}(\mu({x}^*),K(x^*,x^*)) \, \mathrm{d}f \nonumber\\
    &=  \Phi(\mu(x^*),K(x^*,x^*)) . 
\end{align}
The point with the highest probability of improvement is selected. Note that this acquisition function provides a reward \emph{regardless} of the size of improvement.

\subsubsection*{(ii) Expected improvement}
\noindent
The expected improvement improves on the previous result by defining a reward that is dependent on the size of the improvement by defining the utility function:
\begin{equation}
    u(x) = \max( 0, f' - f(x)).
\end{equation}
The acquisition function for the expected improvement is then written as:
\begin{align}
    a_{EI}(x^*) &= \mathbb{E}\left[ u(x^*)| x^*,\mathcal{D} \right] \nonumber\\
    &= \int_{-\infty}^{f'} (f'-f) \mathcal{N}(\mu({x}^*),K(x^*,x^*)) \, \mathrm{d}f  \nonumber\\
    &= [f' - \mu(x^*)]\Phi(\mu(x^*),K(x^*,x^*)) \nonumber\\ 
    &+K(x^*,x^*)\mathcal{N}(\mu(x^*),K(x^*,x^*)) .
\end{align}
The first term is contingent on the size of the improvement, meaning that it will tend to be large for points that are closer to the minimum relative to $f'$. The second term is dependent on the variance of the point $x^*$. Points with large variance will have a large degree of uncertainty, therefore, it makes sense that those points should be explored in order to reduce our uncertainty of the surrogate model. Due to these two terms, this acquisition function encodes a trade-off between \emph{exploitation} due to the first term and \emph{exploration} due to the second term. 

\subsubsection*{(iii) Lower confidence bound}
\begin{figure}[h!]
    \centering
    \includegraphics[width=\columnwidth]{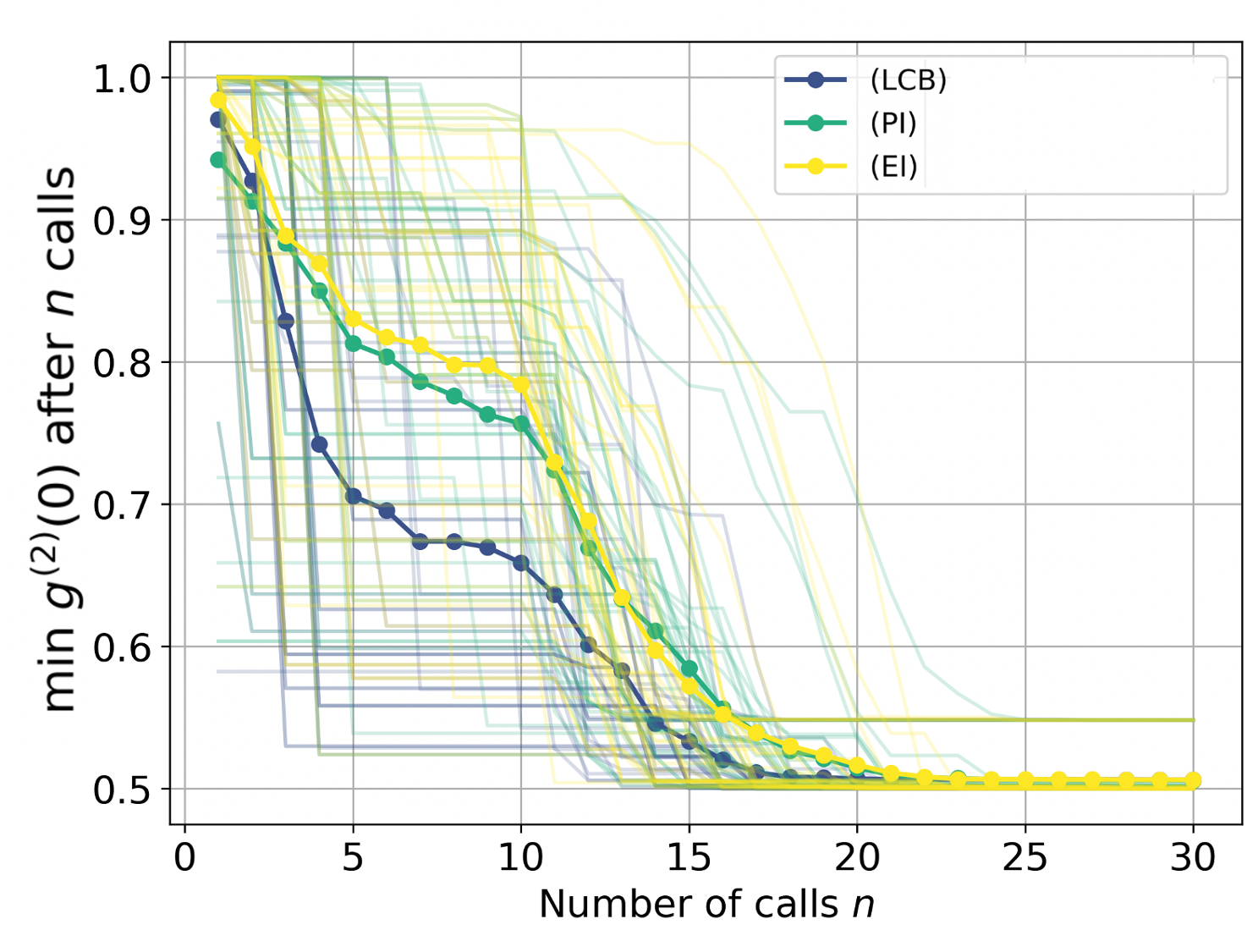}
    \caption{Convergence plots for different acquisition functions where $LCB$ refers to Lower Confidence bound, PI refers to probability of improvement, and EI corresponds to expected improvement.  }
    \label{Acquisition_Comparison}
\end{figure}

\noindent
Here, the lower confidence bound is written as:
\begin{equation}
    a_{LCB}(x^*) = \mu(x^*) - \beta \sigma(x^*),
\end{equation}
where $\beta > 0 $ is a trade-off parameter and $\sigma(x^*)=\sqrt{K(x^*,x^*)}$ is the standard deviation of point $x^*$. Unlike the previous two acquisition functions, this quantity cannot be interpreted in terms of computing the expectation of a utility function, nevertheless, there are strong theoretical results that imply that this acquisition function will converge to the true global minimum of $f$ under certain conditions. A benchmark comparison of different acquisition functions is shown in figure (\ref{Acquisition_Comparison}) demonstrating that the lower confidence bound acquisition function outperforms the other two, where we used $\beta = 2$, as the trade-off parameter value.

\noindent

\subsection{Initial Sampling Scheme Comparison}\label{appsec-inisample}
While Bayesian optimization can start with zero knowledge of the objective function (i.e. the size of training data is null), it is possible to accelerate the convergence of the Bayesian optimization algorithm by using a set of judiciously chosen initial sampling points. Conventionally, there are many techniques available for sampling such as: uniform or random sampling, Sobol sampling, Halton sampling, and Latin hypercube sampling (lhs). By using the $\texttt{skopt}$ initial sampling package, we performed additional benchmarks with respect to the number of initial sampling points as well as the type of sampling used. In Appendix~Fig.~\ref{initial_sampling}, we provide a comparison between Sobol, Halton, random, conventional Latin hypercube sampling and Maximin Latin hypercube sampling showing that the maximin and conventional Lain hypercube sampling provided the best performance. For live testing purposes, we opted to use the conventional Latin Hypercube Sampling scheme, however, further work will be required to determine optimal initial sampling methods for other response functions of interest for quantum network calibration experiments. 
\begin{figure}[h!]
    \centering
    \includegraphics[width=\columnwidth]{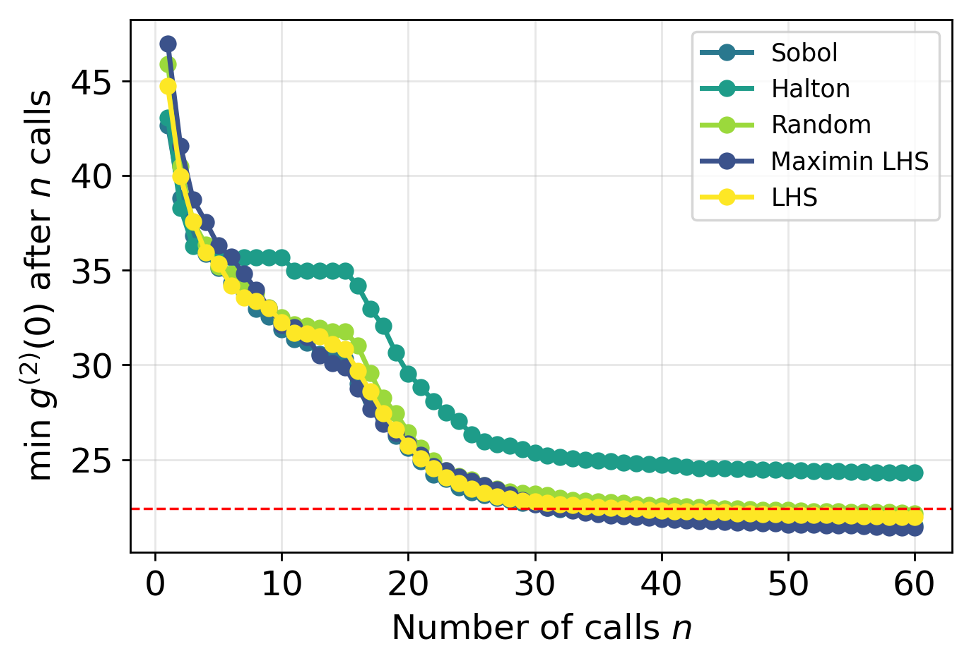}
    \caption{Convergence plots with different initial sampling schemes including Sobol, Halton, random, Maximin Latin Hypercube Sampling, and conventional Latin Hypercube sampling using the the 3D baseline model for testing purposes. }
    \label{initial_sampling}
\end{figure}

\subsection{Summary of Bayesian Optimization Algorithm}\label{appsec-GPoptsummary}

The proposed Bayesian optimization algorithm can be summarized by the following steps:
\begin{enumerate}
    \item Perform $n$ initial measurements $\mathbf{y}=(y_1,\cdots,y_n)$ with respect to the experimental degrees of freedom, $X = [\mathbf{x}_1,\cdots,\mathbf{x}_n]$, using Maximin latin hypercube sampling.
    \item Build Gaussian Process surrogate model based on existing measurements.
    \item Suggest new measurement location $\mathbf{x}_{i+1}$ based on the acquisition function prediction. 
    \item Repeat 2-3 until stopping criterion is met.
\end{enumerate}
The Gaussian Process model built in step 2 of the Bayesian optimization algorithm above can be further decomposed into the following steps:
\begin{enumerate}
    \item Build gram matrix, equation (6), for Matern Kernel ($\nu=5/2$) based on equation (14).
    \item Optimize the Kernel hyperparameters based on the marginal likelihood, equation (17).
    \item Provide predictive mean and variances based on equations (\ref{key1}) and (\ref{key2}).
\end{enumerate}

\subsection{Numerical Implementation}\label{appsec-numimplement}
The numerical implementation and benchmarking of the Bayesian optimization algorithm was done with in-house code written in Python with standard \texttt{numpy} and \texttt{scipy} linear algebra and optimization packages. However, there are a wide variety of excellent open-source software implementations that are available through github. We recommend \texttt{skopt} (Bayesian optimization) and \texttt{GPytorch} (Gaussian Process regression) \cite{Gpytorch} which provide an excellent starting point for initial testing. It is worth noting that we found the customization of the open-source software difficult (for example, adding custom acquisition functions, kernel functions, as well as other functionalities) which is why we opted to perform the benchmarking (shown in this Appendix) with our in-house code. The live testing was performed with a modified version of the \texttt{skopt} Bayesian optimization algorithm. 

\subsection{\textbf{Hong-Ou-Mandel interference model}}\label{appsec-HOMmodel}

To model the experiment we use methods of phase space quantum optics, in particular the characteristic function formalism \cite{LaukInPrep, Takeoka2015}.
This method allows us to describe many experimental imperfections, such as coupling losses, non-perfect detector efficiencies, high photon number contributions, photon number statistics, etc., and is best suited to deal with the Gaussian quantum optical states. Gaussian states are the states whose characteristic function, or in fact any other phase space representation, is given by a multidimensional Gaussian function
\begin{align}
    \chi(\xi)=\exp(-\frac{1}{4}\xi\gamma\xi-id\xi),
\end{align}
with $d$ and $\gamma$ being displacement vector and covariance matrix of the system respectively, $\xi \in \mathbb{R}^{2n}$ with $n$ being the number of independent bosonic mode of the system. The examples of the Gaussian states include the vacuum state, coherent states, as well as squeezed vacuum and two-mode squeezed vacuum states. The later two are directly relevant to our experimental studies since they describe the output states of  degenerate and non-degenerate spontaneous parametric down conversion process respectively.
It is known that linear optic operations, such as beam splitters, phase shifters etc., preserve Gaussian states \cite{Weedbrook2012, Braunstein2005}, i.e. they map Gaussian states onto Gaussian states. For the characteristic function it means that for any linear optic operation there exists a symplectic matrix $S$ that transforms the initial displacement vector and the covariance matrix to the output form, $\gamma^\prime=S^T\gamma S$ and $d^\prime=Sd$.
\begin{figure*}[ht!]
    \center
    \includegraphics[width=0.75\textwidth]{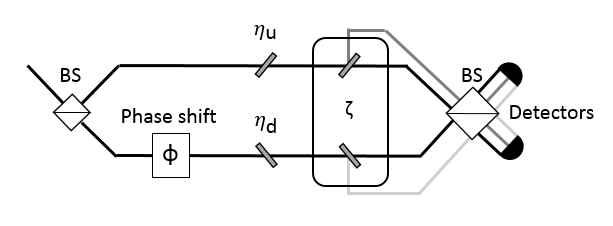}
    \caption{ Representation of the model.  The output of type-0 SPDC crystal, which is described by a squeezed vacuum state, is mixed in with vacuum at a beam splitter.  The coupling efficiencies of the two arms are modeled by a virtual beamsplitters with a transmittance $\eta_u$ and $\eta_d$ for the upper and lower arm respectively. Assuming equal coupling efficiencies we can simulate the polarization match by varying $\eta_u$ from $0$ to $\eta_d$ corresponding to crossed and aligned polarizations respectively. Another virtual beamsplitter with a transmittance $\zeta$ models the mode overlap of the two modes. Only the transmitted parts corresponding to the indistinguishable proportion of the two photons, interfere with each other on the following beamsplitter, while the reflected parts are mixed with vacuum. A controllable phase shift $\phi$ is also present in the bottom arm.}
    \label{fig:app-model}
\end{figure*}

The model representation of the experimental setup is shown in Appendix~Fig.~\ref{fig:app-model}. The output of the SPDC crystal is mixed with a vacuum input on the first beam splitter BS. The overall coupling efficiencies in the upper and the lower arm can be modelled by the virtue of a virtual beam splitter with the transmittance $\eta_u$ and $\eta_d$ respectively. The polarization degree of freedom can be modeled by a variable transmittance, e.g. $\eta_u$, that can take values from zero, corresponding to crossed polarizations, to some finite value $\eta_u^f$, corresponding to parallel polarizations.
Another virtual beam splitter with a transmittance $\zeta$ is used to model the path length or the time of arrival degree of freedom. $\zeta$ corresponds to the mode overlap of the arriving photons and can take values from 0, corresponding to the case of completely distinguishable photon wave packets with zero overlap, to 1, corresponding to the case of completely indistinguishable photons. To resemble the experimental data more closely we parametrize the $\zeta$ parameter as a Gaussian function $\zeta=\exp(-x^2)$ reflecting the fact that the overlap of two Gaussian pulses is again a Gaussian function.  As we can see from the figure only transmitted parts interfere at the following beam splitter whereas reflected parts do not interfere but are mixed with a vacuum input instead.
To model the phase averaging we introduce a phase shifter in the lower arm that shifts the relative phase between the two arms by an amount $\phi$. To obtain the experimental values of the (phase dependent) quantities we average over $\phi$ from $0$ to $2\pi$. In the experimental setup this averaging is achieved by the phase-modulator in one of the arms continuously sweeping over a $2\pi$ phase-shift.

After determining individual symplectic transformations and combining them together we obtain the covariance matrix that completely describes the final state. Using the characteristic function description we can now calculate the probabilities for a single and coincidence photon detection as follows:
\begin{align}
    P_{D1/D2}&=\mathrm{Tr}\{\hat{\rho}(1-\ket{0}\bra{0}_{1/2})\}=1-\mathrm{Tr}\{\hat{\rho}(\ket{0}\bra{0})_{1/2}\}\\
    P_{D1D2}&=\mathrm{Tr}\{\hat{\rho}(1-\ket{0}\bra{0}_1)(1-\ket{0}\bra{0}_2)\} \nonumber \\
    &= 1-\mathrm{Tr}\{\hat{\rho}(\ket{0}\bra{0})_{1}\}-\mathrm{Tr}\{\hat{\rho}(\ket{0}\bra{0})_{2}\} \nonumber \\
    &+ \mathrm{Tr}\{\hat{\rho}(\ket{0}\bra{0})_{1}(\ket{0}\bra{0})_{2}\},
\end{align}
where $\ket{0}_{1/2}$ corresponds to a vacuum state of all modes impinging on the detector $D1/D2$.
From these we can now easily determine the normalized $g^{(2)}(0)$ correlation function as:
\begin{align}
    g^{(2)}(0)=\frac{P_{D1D2}}{P_{D1}P_{D2}}.
\end{align}

We studied the $g^{(2)}(0)$ correlation function for different input states corresponding to different physical situations. If the underlying nonlinear process is a degenerate down conversion, i.e. two identical photons are created, the input state is described by a squeezed vacuum state with the corresponding covariance matrix
\begin{align}
    \gamma_{SV} = \begin{pmatrix}
    \mathrm{e}^{-2r} & 0\\
    0 & \mathrm{e}^{2r}
    \end{pmatrix},
\end{align}
where $r$ is the squeezing parameter. This input corresponds to all of the experimental two dimensional parameter space scans, i.e., the data in Fig.~\ref{map} of the main text. In Appendix~Fig.~\ref{fig:app-g2model}a the $g^{(2)}$ is plotted as a function of $\zeta=\exp(-x^2)$ (path length difference) and $\eta_u$ (loss due to polarization misalignment of upper arm).

\begin{figure*}[t]
    \centering
    \includegraphics[width=\textwidth]{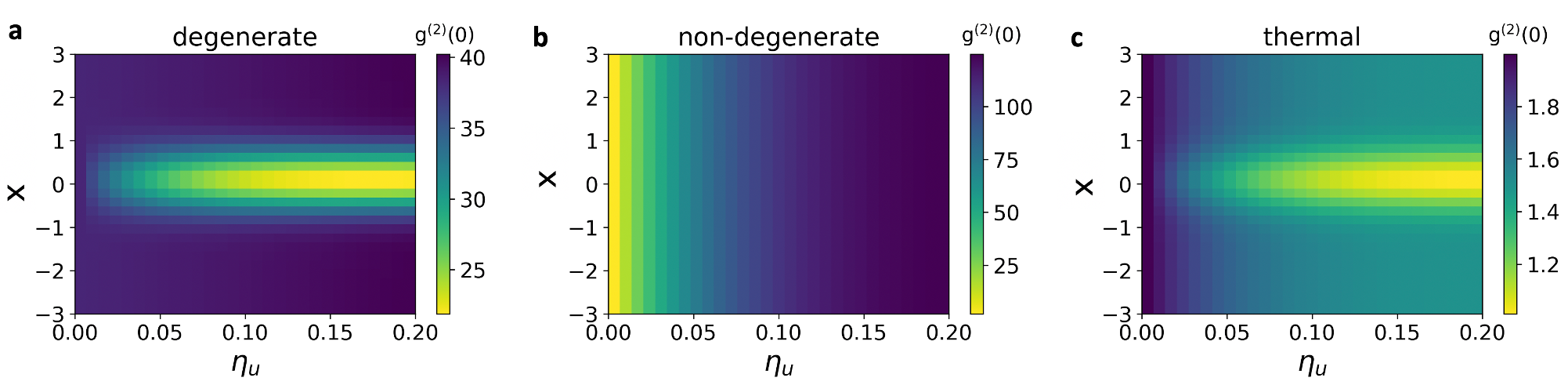}
    \caption{Normalized $g^{(2)}(0)$ correlation function for (a) degenerate single-mode squeezed vacuum input state, and (b) non-degenerate two-mode squeezed vacuum input state, and (c) two thermal input states. }
    \label{fig:app-g2model}
\end{figure*}

If the down conversion process is non-degenerate, i.e. two different photons are created for example at different frequencies, the input state is a two-mode squeezed vacuum state with the corresponding covariance matrix
\begin{widetext}
\begin{align}
    \gamma_{TMSV}= 
    \begin{pmatrix}
    \cosh^2(r)+\sinh^2(r) & 0 & 2\cosh(r)\sinh(r) & 0 \\
    0 &  \cosh^2(r)+\sinh^2(r) & 0 & -2\cosh(r)\sinh(r)\\
    2\cosh(r)\sinh(r) & 0 & \cosh^2(r)+\sinh^2(r) & 0 \\
    0 & -2\cosh(r)\sinh(r) & 0 & \cosh^2(r)+\sinh^2(r)
    \end{pmatrix}
\end{align}
\end{widetext}

This input corresponds to the case in the three-dimensional parameter space for which the spectral filters in the two arms have been shifted with respect to each other by an amount much greater than the bandwidth of the filters. However, in this case the photons in each arm are clearly distinguishable, so the final HOM interference will always be between distinguishable modes. We can tweak the model to accommodate this feature by introducing another virtual beamsplitter with reflectivity $\zeta'$. Essentially, the same result can be achieved by fixing $\zeta=0$. The predicted  $g^{(2)}(0)$ is plotted as a function of $\zeta=\exp(-x^2)$ (path length difference and $\eta_u$ (loss due to polarization misalignment of upper arm in Appendix~Fig.~\ref{fig:app-g2model}b. The cross-correlation function does not depend on $\zeta$, but only on $\eta_u$, since the latter induces actual loss in the PBS, i.e., changes the number of photons reaching the detector. In reality we do not fully achieve the scenario in which the filters are detuned by more than bandwidth as the filters are maximally shifted by 6~GHz with respect to each other while their bandwidth is about 12~GHz. Hence, the actual expected $g^{(2)}$ will be a combination of the squeezed vacuum input case (Appendix~Fig.~\ref{fig:app-g2model}a) and the non-degenerate two-mode squeezed state (Appendix~Fig.~\ref{fig:app-g2model}b). The relative weight of the two contributions can be found by calculating the spectral overlap of the photons after the filter, which, for Gaussian filter pass-bands, will again yeild a Gaussian shaped weight parameter as a function of the spectral separation of the two filters.

Finally, in a real world implementation of a quantum network the Bell state measurements will occur between two different entangled photon sources. This means that the two partaking photons will originate from different SPDC sources and, thus, not be correlated. The input state in this case will correspond to two thermal states with the corresponding covariance matrix 
\begin{align}
    \gamma_{TMSV}=\begin{pmatrix}
    1+2\mu & 0 & 0 & 0 \\
    0 &  1+2\mu & 0 & 0\\
  0 & 0 & 1+2\mu^\prime & 0 \\
    0 & 0 & 0 & 1+2\mu^\prime
    \end{pmatrix},
\end{align}
where $\mu$ and $\mu^\prime$ are the mean photon numbers at the two inputs. Appendix~Fig.~\ref{fig:app-g2model}c shows the $g^{(2)}(0)$ function for the thermal input state scenario.

\end{document}